\documentclass[journal]{IEEEtran}
\usepackage[utf8]{inputenc}
\usepackage[T1]{fontenc}
\usepackage{xcolor,cite,etoolbox}
\usepackage{color,soul}
\usepackage{amsmath}
\usepackage{amssymb}
\usepackage{flushend}
\usepackage{url}
\usepackage{graphicx}
\usepackage{caption}
\definecolor{aqua}{rgb}{0.0, 1.0, 1.0}
\definecolor{babyblue}{rgb}{0.54, 0.81, 0.94}
\usepackage{caption}
\usepackage{subcaption}
\usepackage{lipsum}
\ifCLASSINFOpdf
\else
\fi
\begin{document}
	%
	\title{Digital RIS (DRIS): The Future of Digital Beam Management in RIS-Assisted OWC Systems}
	%
	%
	%
	
\author{Alain R. Ndjiongue, Telex M. N. Ngatched, Octavia A. Dobre, and Harald Haas 
\noindent
\thanks{A. R. Ndjiongue, T. M. N. Ngatched, and O. A. Dobre are with the Faculty of Engineering and Applied Science, Memorial University, Canada.\\ Harald Haas is with the LiFi Research and Development Center, Dpt. Electronic and Electrical Engineering, the University of Strathclyde, Glasgow, U.K.}}
	\maketitle
	
	\begin{abstract}
Reconfigurable intelligent surfaces (RIS) have been recently introduced to optical wireless communication (OWC) networks to resolve skip areas and improve the signal-to-noise ratio at the user's end. In OWC networks, RIS are based on mirrors or metasurfaces. Metasurfaces have evolved significantly over the last few years. As a result, coding, digital, programmable, and information metamaterials have been developed. The advantage of these materials is that they can enable digital signal processing (DSP) techniques. For the first time, this paper proposes the use of digital RIS (DRIS) in OWC systems. We discuss the concept of DRIS and the application of DSP methods to the physical material. In addition, we examine metamaterials for optical DRIS with liquid crystals serving as the front row material. Finally, we present a design example and discuss future research directions.
	\end{abstract}
	

	\IEEEpeerreviewmaketitle
	
\section{Introduction} \label{intro}
Due to the high bandwidth demands of network users, the actual implementation of beyond fifth-generation (B5G) mobile networks may also rely on light waves for transmission within indoor environments. Signals at these frequencies suffer high propagation losses, which can be effectively mitigated by using beamforming techniques and multiple-input multiple-output (MIMO). These techniques aim to improve the system's spectral efficiency and provide a cost-effective solution with reliable coverage \cite{8454520}. There is the possibility that B5G mobile networks may still be susceptible to signal loss due to obstructions, resulting in the skip-zone problem. In recent studies, reconfigurable intelligent surfaces (RIS) were incorporated into optical wireless communication (OWC) networks as an effective tool to resolve this dilemma and increase the signal-to-noise ratio (SNR) at the user's end \cite{9614037, 9443170, 9276478, 9466323, 9525295, 9578932}. Moreover, RIS can improve beamforming gains compared with massive MIMO, enhance spectral and energy efficiency, and improve network coverage, to name a few more advantages \cite{9525295, 8811733, 9133435}.

In OWC, beamforming is accomplished by using multiple light sources (LSs) to send the same signal with equal wavelengths (and phases). Therefore, the targeted light beam is provided with enhanced signal strength in a specific direction. As the number of LSs increases, the beamwidth becomes narrower. By adjusting the lighting parameters, these LSs can also perform beam steering in addition to beamforming. Thus, the transmitter can simultaneously target multiple users with different beams, whose characteristics may change as the users move in separate directions. Furthermore, a multiple LSs transmitter may switch beams if the actual beam is unable to track the corresponding user. In contrast with wireless fidelity, which generally utilizes MIMO techniques, light fidelity will also rely on techniques such as spatial diversity at the layer switches along with powerful digital signal processing (DSP) techniques.

Metasurfaces have experienced rapid development recently, propelling them to the forefront of RIS materials. Consequently, the RIS technology has been integrated into the wireless network to solve skip zones and improve the SNR at the receiver. Subsequently, the RIS has been analyzed for free-space optical and visible light communications (VLC) systems \cite{9525295, 9578932}. The RIS element transmits the incident signal in a unique direction and intensity depending on its phase shift and transition coefficient. In contrast to single-element RIS, multiple-element RIS are excellent at focusing the signal at specific users. In addition, efficient algorithms can be applied to manipulate these RIS elements. Considering that metamaterials are regarded as analog materials of optical RIS, most elements are composed of artificially layered metamaterial structures, which can be seen as analog materials due to the continuous nature of the medium \cite{cui2014coding, cui2017information}. Recent developments in the metasurface technology have revealed more advantages, including the ability to digitally control the phase and amplitude of propagation waves. The coding metasurfaces, digital metasurfaces, and programmable metasurfaces are digitized forms of metasurfaces \cite{cui2014coding, cui2017information}. Further, space-time-coding (STC) digital metasurfaces have been proposed in \cite{zhang2018space}, with significant advantages for wireless communications and radar systems. These digital metasurfaces form the basis of the digital RIS (DRIS), which is proposed in this paper.  

In the proposed design, a software program controls the processes of scattering, absorption, reflection, diffraction, and refraction in RIS elements. In the resulting DRIS, a digital code is assigned to each element, allowing the physical materials to be viewed as binary sequences. DRIS will result in improved control and management of the transmission environment. In addition, DSP techniques, such as STC, can be seamlessly implemented over the physical material \cite{zhang2018space}. A number of advantages are offered by STC, including spatial diversity, better performance in fading environments, minimal decoding complexity, and maximum diversity gain, among others. This article introduces the DRIS technology, explains how to perform DSP techniques within the OWC transmission environments, and describes its structural implementation. The design presented in this paper offers a powerful solution to perform and enhance beamforming in RIS-assisted OWC systems. The DRIS elements are characterized by a sequence of bits "0" or "1", a reflection coefficient, and a phase shift. The main advantage of DRIS over its analog counterpart is its ability to enable DSP techniques such as STC, for example, into physical materials.  
\begin{figure}
	\centering
	\includegraphics[width=0.4\textwidth]{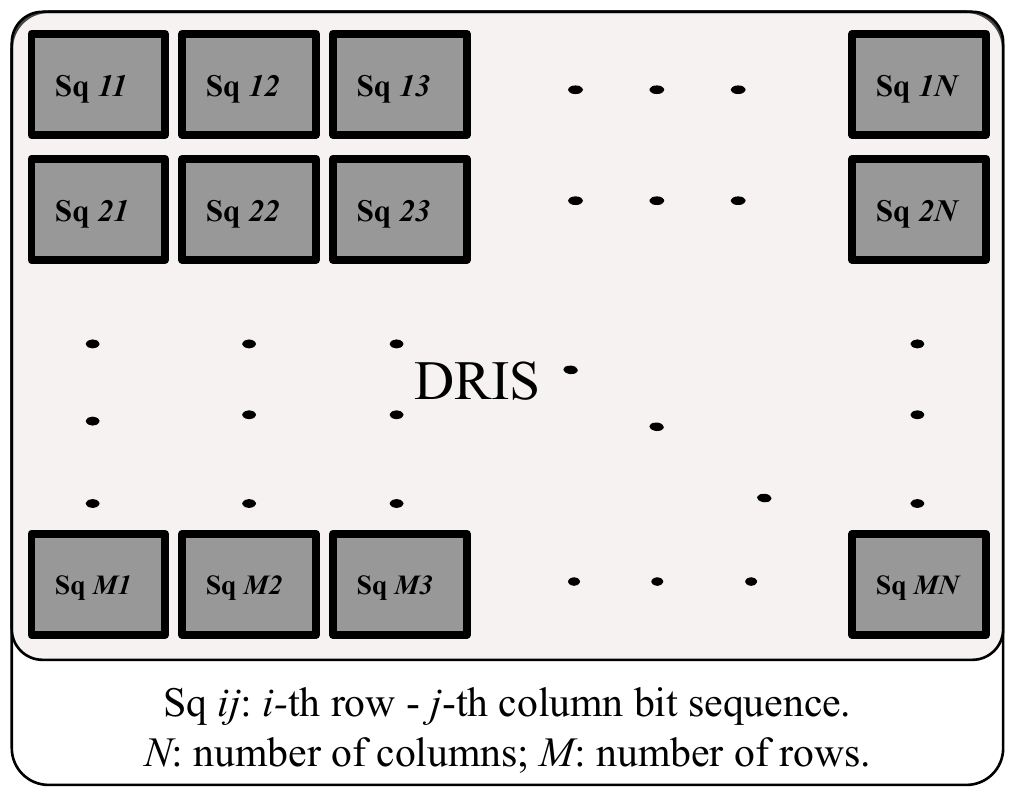}
	\caption{The concept of digital RIS (DRIS).}
	\label{fig:DRIS}
\end{figure}
\begin{table}
	\caption{Examples of 1-bit, 2-bit, and 3-bit DRIS coding. \label{Table1}}
		\centering
	\begin{tabular}{l|c|c|c|c} 
	\hline \hline
	Phase shift (rds)& 0 & $ \pi/2 $ & $  \pi $& 3$ \pi/2 $ \\ \hline
	Reflection/refraction coefficient& $\lambda_1$ & $ \lambda_3 $ & $\lambda_2$& $\lambda_4$\\ \hline
	1-bit & 0 &//  &1 & //  \\ 
	2-bit & 00 &01 & 11& 10  \\ 
	\hline \hline
\end{tabular}
\end{table}
\section{The Concept of Digital RIS (DRIS)} \label{Concept}
Coding metamaterials are made up of elements that contain a series of "0"s and "1"s. On the other hand, the digital metamaterial is described as a metasurface in which the elements contain either a "0" or "1" \cite{Ref_Light}, while a programmable metamaterial is described as a metasurface in which the "0" and "1" states of the elements are controlled by a programmable device. Metamaterials can be digitally manipulated in order to introduce signal processing methods similar to those used in communication theory into the material. The concept is known as information metamaterial \cite{Ref-Chemist}. As a result, the authors in \cite{zhang2018space} have introduced and presented STC digital metasurfaces. It should be noted that many other processing algorithms can similarly be incorporated into the physical material, with multiple applications in wireless communications, radars, and imaging. These theories of digital and programmable metamaterial are the foundation of DRIS.
\subsection{The Concept}
DRIS are RIS to which sequences of bits are transmitted in order to control individual elements. Specially, a DRIS is a RIS defined by the code ($X$, $ k $, $\rho_o$), where $X = M \times N$ represents the number of RIS elements, $ k = \log_2(L)$, $ L $ being the DRIS level, indicating the number of element types in the RIS, and $\rho_o$ is the initial transition coefficient (the $u$-indexed RIS transition coefficient is defined by $\Gamma_u = \beta_u\rho_o$, where $\beta_u$ is the $ u $-indexed coefficient). It should be noted that $ u $ is an index that describes both the emerging light orientation and its power. The concept is illustrated in Fig.~\ref{fig:DRIS} and Table~\ref{Table1}. Let $\theta$ be the angle of reflection/refraction, $\theta = 2u\pi/L$, where $ u $ = \{1, 2, 3, 4, \dots, $ L $\}. For example, the DRIS (1000, 2, 0.75) is a 2 bits, 1000 elements DRIS, with $ u \in $ \{1, 2, 3, 4\}, on which the incident light will be reflected (refracted) into four different orientations (0, $\pi$/2, $\pi$, 3$\pi$/2), with four different amplitudes 0.75$\beta_1$, 0.75$\beta_2$, 0.75$\beta_3$, and 0.75$\beta_4$, where $\beta \in \mathbb{N}^*$, is defined based on the design requirements. 
\subsection{DRIS Coding}
DRIS is coded assuming a power-of-two number of bits per DRIS element. As an example, in a binary DRIS, each element is represented by a single bit, "0" or "1". It may be difficult in practice to maintain the phase difference between the bits "0" and "1" since this is related to the frequency of the incoming signal. The major difference between traditional RIS and DRIS lies in the way the metasurface controls the incoming light. The traditional RIS uses the metasurface parameters to manipulate the signal, while the DRIS utilizes a coding sequence to steer and attenuate/amplify the light wave. A DRIS ($ X $, 1, $ \rho_o $) uses two types of elements, while a DRIS ($ X $, 2, $ \rho_o $) uses four types of elements. In general, a DRIS ($ X $, $ k $, $ \rho_o $) has $ 2^k $ ($ 2^k \leqslant X$) types of elements. In order for the algorithm used to control the DRIS to work, it must be able to switch codes between elements of the DRIS to allow the beam formed by the reflected light to shift and trace the user as it moves.
\begin{figure*}
	\centering
	\includegraphics[width=0.9\textwidth]{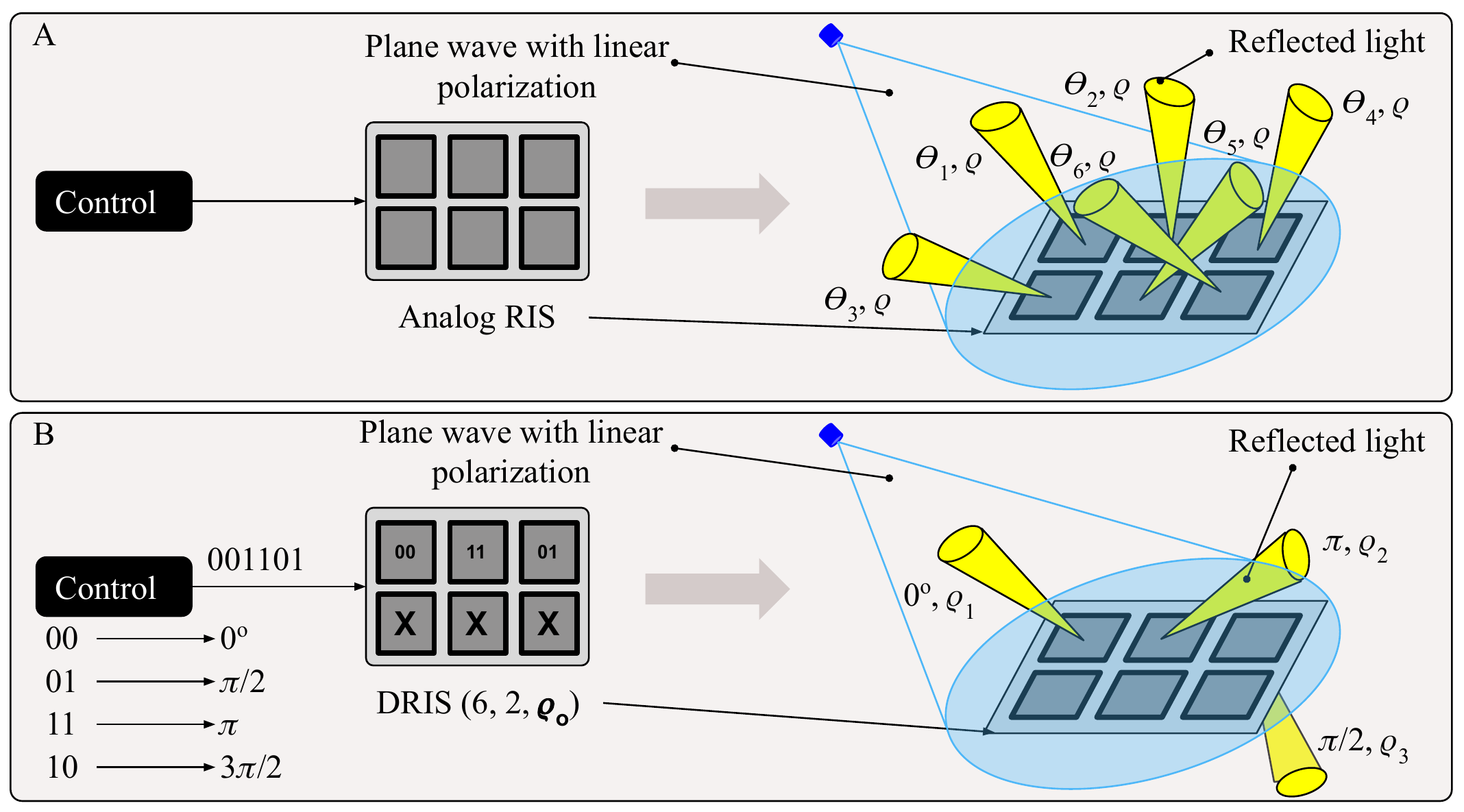}
	\caption{Analog RIS vs. DRIS: A- Analog RIS with random element orientation; B- 2-bit coding RIS.}
	\label{fig:Focus}
\end{figure*}
\begin{figure*}
	\centering
	\includegraphics[width=0.9\textwidth]{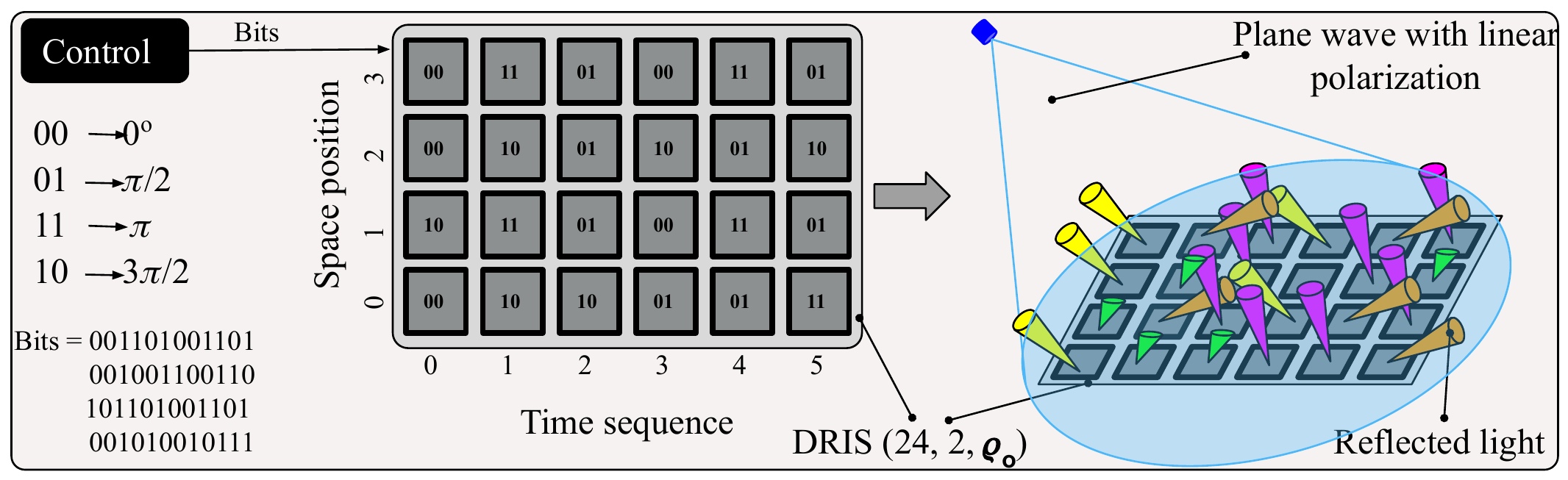}
	\caption{Space-time coded DRIS ($ X $, 2, $\rho_o$).}
	\label{fig:Space_Time_Coding}
\end{figure*}

Fig.~\ref{fig:Focus} compares the traditional RIS and DRIS. In this figure, the RIS is made of a number of similar elements, which are controlled numerically. All elements are controlled individually with respect to their phase shifts. Fig.~\ref{fig:Focus}-B illustrates a DRIS ($ X $, 2, $\rho_o$). The DRIS module consists of four types of elements (Type 1, Type 2, Type 3, and Type 4). Each of these elements is controlled numerically with the bit sequences "00", "01", "11", and "10", respectively. Consequently, a DRIS ($ X $, 2, $\rho_o$) consists of multiple elements of Types 1 to 4. Accordingly, the phase shifts are 0, $\pi$/2, $\pi$, and 3$\pi$/2. The DRIS elements could be of any of the four types depending on the received control sequence. As an example, consider a DRIS ($ X $, 2, $\rho_o$), where the incoming light falls with a unique incidence angle. The digital control sequence "00000000" only engages one type of element, which deflects the incoming light with a 0$ ^o $ shift. For the digital control code sequence 01000111, the DRIS consists of three types of elements. As a result, the reflected light beams will have three orientations, $\pi$/2, 0, and $\pi$. Note that the control bit sequence determines the properties of the elements, which can result in either reflection, refraction, absorption, or filtering effects. 
\subsection{Space-Time-Coding (STC) DRIS}
Due to its digital nature, DRIS can enable DSP techniques such as STC. The resulting structure, which we refer to as STC-DRIS, can improve the performance and the reliability of data transmission through the DRIS, and therefore the performance of RIS-assisted OWC systems. The STC-DRIS is illustrated in Fig.~\ref{fig:Space_Time_Coding}. A control bit sequence is grouped into blocks of bits based on the DRIS resolution (1-bit, 2-bit, etc.) and transmitted sequentially to the DRIS. In the first time slot, the first bit sequence is allocated to the DRIS elements in their respective space positions. Other bit sequences are sent to the DRIS in the next time slots and allocated to the corresponding space positions. The same pattern is followed to fill up the remaining spaces. For a specific period of time, the transmitted sequence determines the characteristic of the element. Subsequently, these control sequences determine the element's behavior (for example, refraction or reflection) and decide the direction and intensity of the emerging lights. It should be noted that STC is only an example of a DSP method that can be applied to the DRIS. STC coding and any other DSP  method/algorithm can be implemented over tunable physical metamaterials. Thus, DRIS has the potential to revolutionize the world of communication in the area of controllable communication media. 
\section{Optical-Based DRIS Design} \label{design}
\subsection{DRIS Metamaterials}
\begin{figure}
	\centering
	\includegraphics[width=0.4\textwidth]{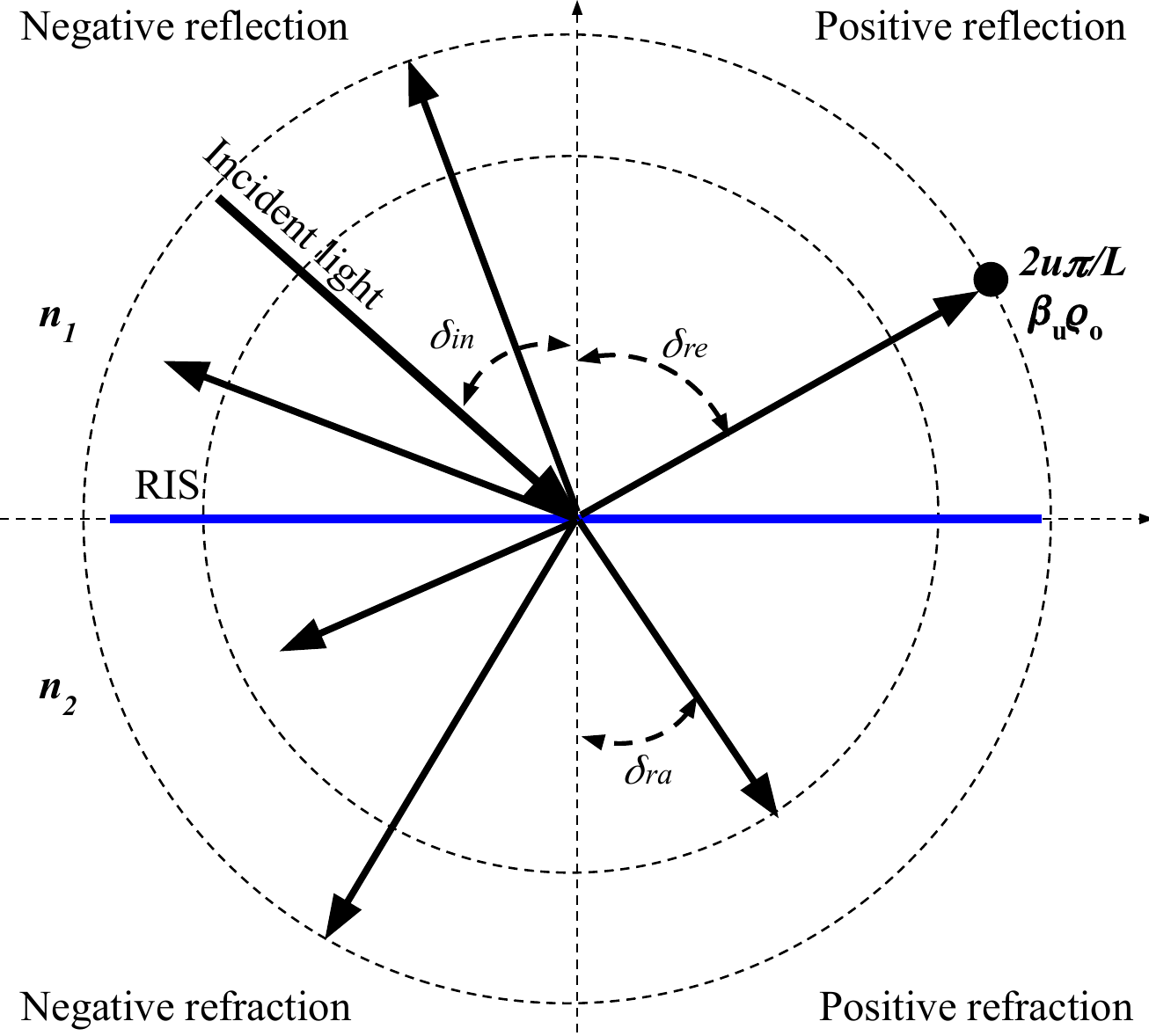}
	\caption{Positive and negative reflection/refraction.}
	\label{fig:Number}
\end{figure}

Consider two ordinary media with refractive indices $ n_1 $ and $ n_2 $, respectively. The reflection and refraction phenomena that can occur when an incident light travels from the first medium to the second are summarized in Fig.~\ref{fig:Number}. The corresponding modification of the Snell's law is given by \cite{book3}
\begin{equation}
	\left\{\begin{array}{@{}l@{}}
		n_2\sin(\theta_{ra}) = n_1\sin(\theta_i) + \frac{q}{\chi}\\
		n_1\sin(\theta_{re}) = n_1\sin(\theta_{i}) + \frac{q}{\chi}
	\end{array}\right.\,,
\end{equation}
where $\theta_{i}$, $\theta_{re}$, and $\theta_{ra}$ are respectively the angles of incidence, reflection, and refraction, $\chi = \phi (x, y)/[n(x, y)d] $ is the wavevector along reflection or refraction axis, with $ d $ as the thickness of the DRIS environment of index $ n_2 $, $\phi (x, y)$ as the phase retardation, and $ q $ = $\partial(\Delta \phi)/\partial(x,y)$, with $ \partial(\cdot) $ as the partial derivative operation. For a phase shift of 2$\pi$, it is necessary to have $ d $ = $\lambda$, where $\lambda$ is the wavelength of the incident light. According to the value and sign of $ q $, the emerging light occurs in one of the four quadrants shown in Fig.~\ref{fig:Number}. Traditional materials can be used to obtain positive reflection and refraction. For negative reflection and refraction, special materials may be required. Some metamaterials that can be used in these types of systems are metallic nanospheres in a dielectric medium, thin metallic rods isotropically distributed in a dielectric medium, split-ring metallic elements in a dielectric medium, negative-index metamaterials, hyperbolic metamaterials, metasurfaces, complementary metasurfaces, and liquid crystals (LCs)-based metasurfaces \cite{book3, 6167595, shalaev2007optical, wang2007tunable}. A metasurface is controlled by its geometric structure and its material parameters \cite{9474926}. For soft materials such as liquid crystals, metal fluids, liquid metals, and bio-materials, the primary material parameter is the refraction index \cite{9354893}. In addition to this parameter, the amount of light generated also depends on the LC substance and dye, as well as their concentration. Compression and expansion of solid materials, such as elastomers, may be useful for modifying the metasurface structure, which, according to \cite{9474926}, influences both the direction and amplitude of incoming light. It should be noted that elastomers need physical variations in thickness in order to change their optical parameters \cite{9474926}. Although this approach is effective, it may result in the deterioration of the elastomer. As an alternative, a DRIS containing LC-based elements can be designed. As they are capable of steering a light with voltage-controlled reflectance and transmission, LCs are appropriate materials for use in DRIS devices when placed between two polarizers \cite{9354893}. This ability can be measured by varying the externally applied voltage, which will allow for complete control of both the DRIS phase shift and transition coefficient.
\subsection{LC-based DRIS}
There has been a trend toward the use of LCs as front-row materials for light beam and intensity manipulators, optical phased arrays, and beam steerers \cite{9354893, 1035273, 4939409, willekens2016reflective}. LCs embedded in a layered structure are able to control reflection and refraction of light at a specific temperature of the cavity \cite{ndjiongue2021design}.

\noindent
\textbf{LCs' tuning:} At a given temperature, an element's phase shifter can be tuned by varying the LC thickness, $ d $, birefringence, $\Delta n$, or changing the light color, $\lambda$. The value of $d$ is determined during the fabrication of the RIS element and the value of $\lambda$ is independent of the structure of the element. Therefore, the retardation angle, $\phi$ (deflection/refraction), depends only on the birefringence, $\Delta n$. As demonstrated in \cite{9354893}, the birefringence is sensitive to an incoming electric field variation. LC-based DRIS elements can be controlled by an external electric field in two ways: ($ i $) all elements are exposed to the same external electric field. Accordingly, each element type possesses a different LC composition, which may be the same LC substance coloured with a different dye; ($ ii $) each element is exposed to its own external electric field. In this case, all elements are subjected to the same LC composition. The second option would undoubtedly lead to a higher level of design complexity and manufacturing costs. However, it would present an improved degree of tuning flexibility.

\noindent
\textbf{LCs' refractive indices:} LCs possess two different refractive indices as a result of their anisotropic nature, namely the ordinary index, $ n_o $, and extraordinary index, $ n_e $. In addition, they are characterized by an optical axis, which is the line around which a rotation will not alter the optical properties of the material. The propagation of an optical wave in the LC is divided into two components, an ordinary ray along the ordinary axis governed by $ n_o $, and an extraordinary ray along the extraordinary axis parallel to the optical axis and governed by $ n_e $. In practice, the LC birefringence, $\Delta n = n_0 - n_e $, is determined by the orientation-dependent differences in the refractive indices of the LC. Upon entering the LC, the light is refracted by the substance's refractive index. The light is retarded as a result of this change in velocity as it passes through the LC substance. Higher birefringence LC substances can provide larger retardation angles. Therefore, the deflected or diffracted light signal is a measure of the dynamic variation of $\Delta n$.

\begin{figure}
	\centering
	\includegraphics[width=0.45\textwidth]{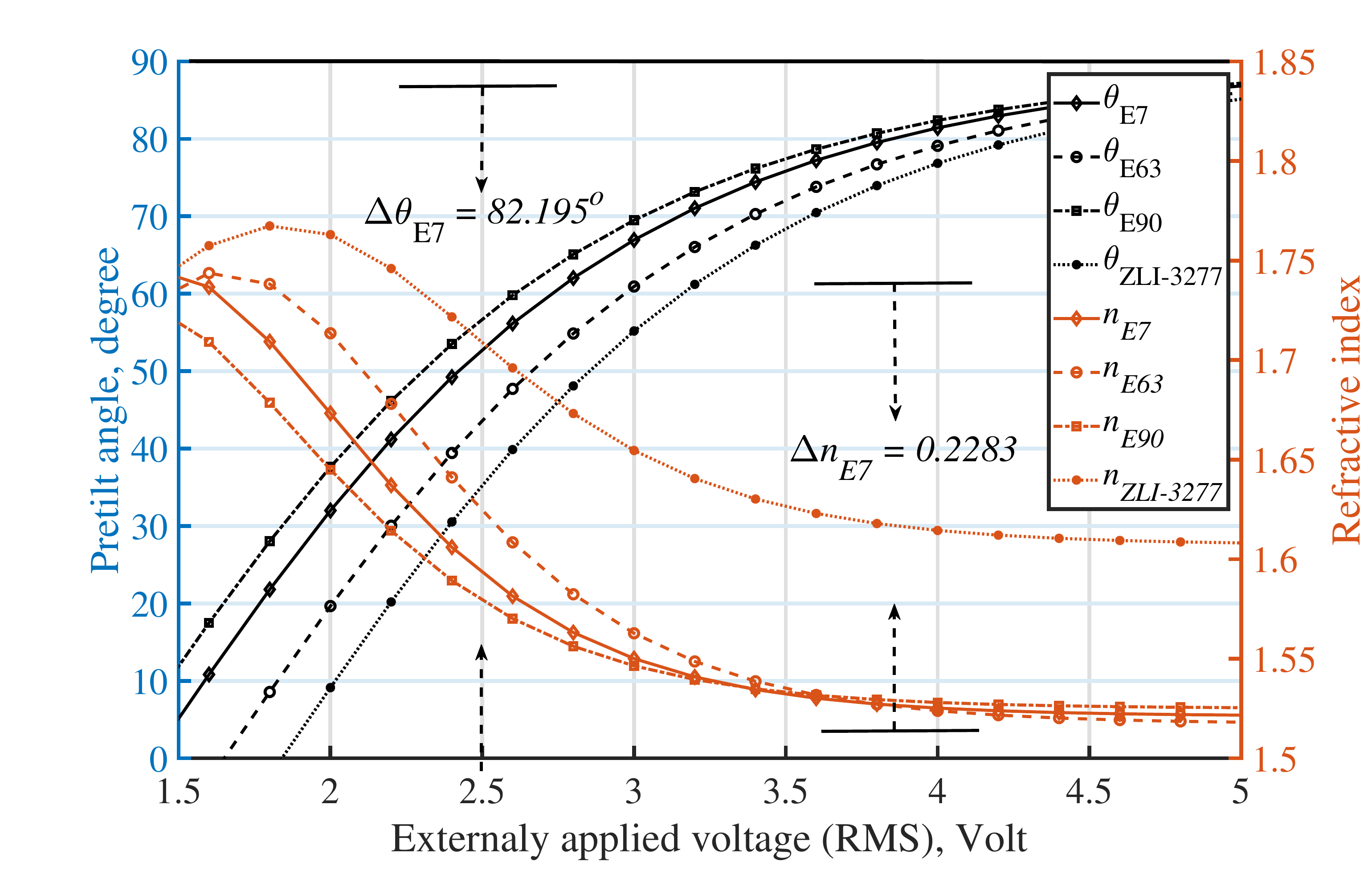}
	\caption{Tilt angles and refractive index.}
	\label{fig:Tilt}
\end{figure} 
Tuning an LC involves varying the tilt angle, $\psi$, between the crystal orientation and the LC's optical axis. Due to this variation, the refractive index is given by \cite{book3}
\begin{equation}
	\frac{1}{n^2(\psi)} = \frac{\cos^2(\psi)}{n_e^2(\psi)} + \frac{\sin^2(\psi)}{n_o^2(\psi)}.
	\label{Eq:2}
\end{equation}
The refractive index directly influences the retardation angle, $ \phi $, whose maximum value is $\phi_{max} = 2\pi d \Delta n/\lambda$. However, the retardation angle varies with the refractive index in \eqref{Eq:2} as $\phi(\psi) = 2\pi[n(\psi) - n_o]/\lambda$. This shift can also be measured through the normalized retardation, $\phi(\psi)/\phi_{max}$ = $(n(\psi) - n_o)/(n_e - n_o)$. A very effective and smooth method of controlling the refractive index is to apply an externally controlled electric field to the LC. Its impact on the tilt angle, $\psi$, can be modeled as \cite[Eq. (21.3-1)]{book3}
\begin{equation}
\psi = \left\{\begin{array}{@{}l@{}}
		0, \:\:\: V_c > V \\
		\frac{\pi}{2} - 2\tan^{-1} \left[\exp(V_c - V)\right], \:\:\: V > V_c,
	\end{array}\right.\,
\label{Eq:3}
\end{equation}
where $ V $ and $ V_c $ are respectively the externally applied voltage and the LC threshold voltage. It is pertinent to note that $\Delta n$ is dependent on the environmental temperature, the external electric field, and the LC substance. In practice, there are several types of LC available for use. These may vary depending on the crystal shape. As an example, three types of thermotropic LC are identified, namely discotic, conic, and rod-shaped. The order parameters of these LCs are highly dependent on the element temperature. These LCs are classified by their phase (nematic, smectic, and discotic), which is a state that exists between the solid and liquid states. Further information on LCs can be found in \cite{book1, book2}. An example of LC tilt angle and refractive index control, generated from the models in \eqref{Eq:2} and \eqref{Eq:3}, is provided in Fig.~\ref{fig:Tilt} for the commercially available E7, E63, E90, and ZLI-3277 LC samples, at room temperature. E7 can exhibit a variation of up to 0.46$\pi$, corresponding to a variation of about 0.2283 in the refractive index. It should be noted that doping LC with a dye may dramatically affect $\psi$, $n $, $\phi$. 
\begin{figure}
	\centering
	\includegraphics[width=0.48\textwidth]{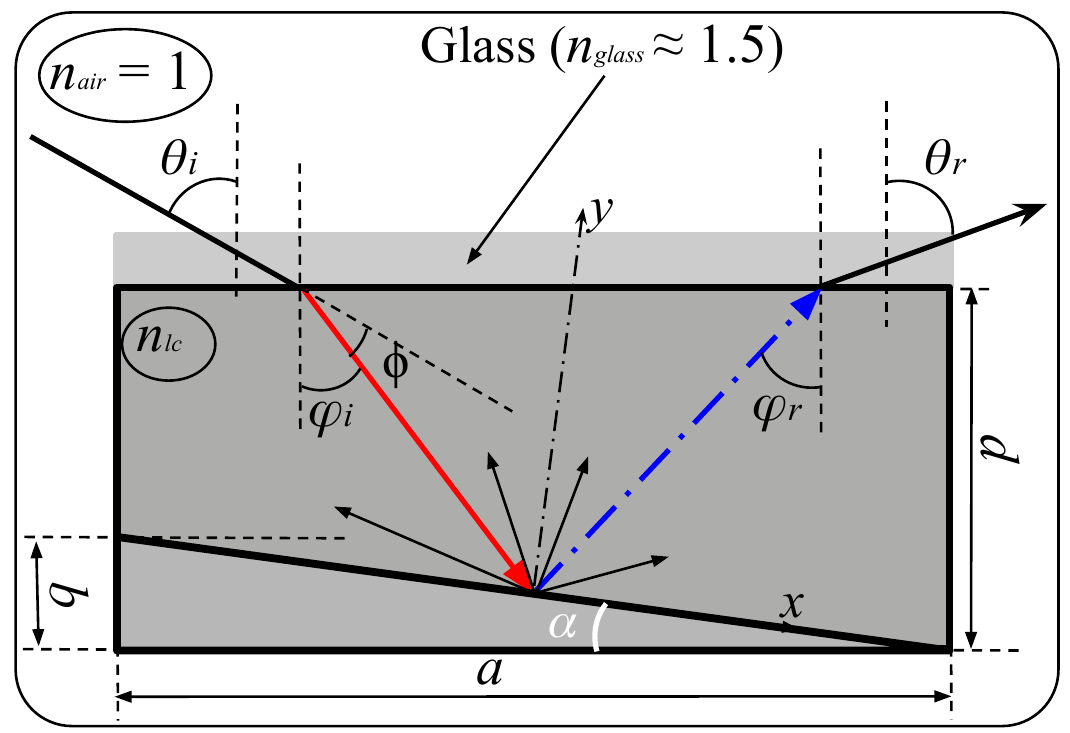}
	\caption{DRIS geometric optics.}
	\label{fig:Design}
\end{figure}  
\begin{figure}
	\centering
	\includegraphics[width=0.5\textwidth]{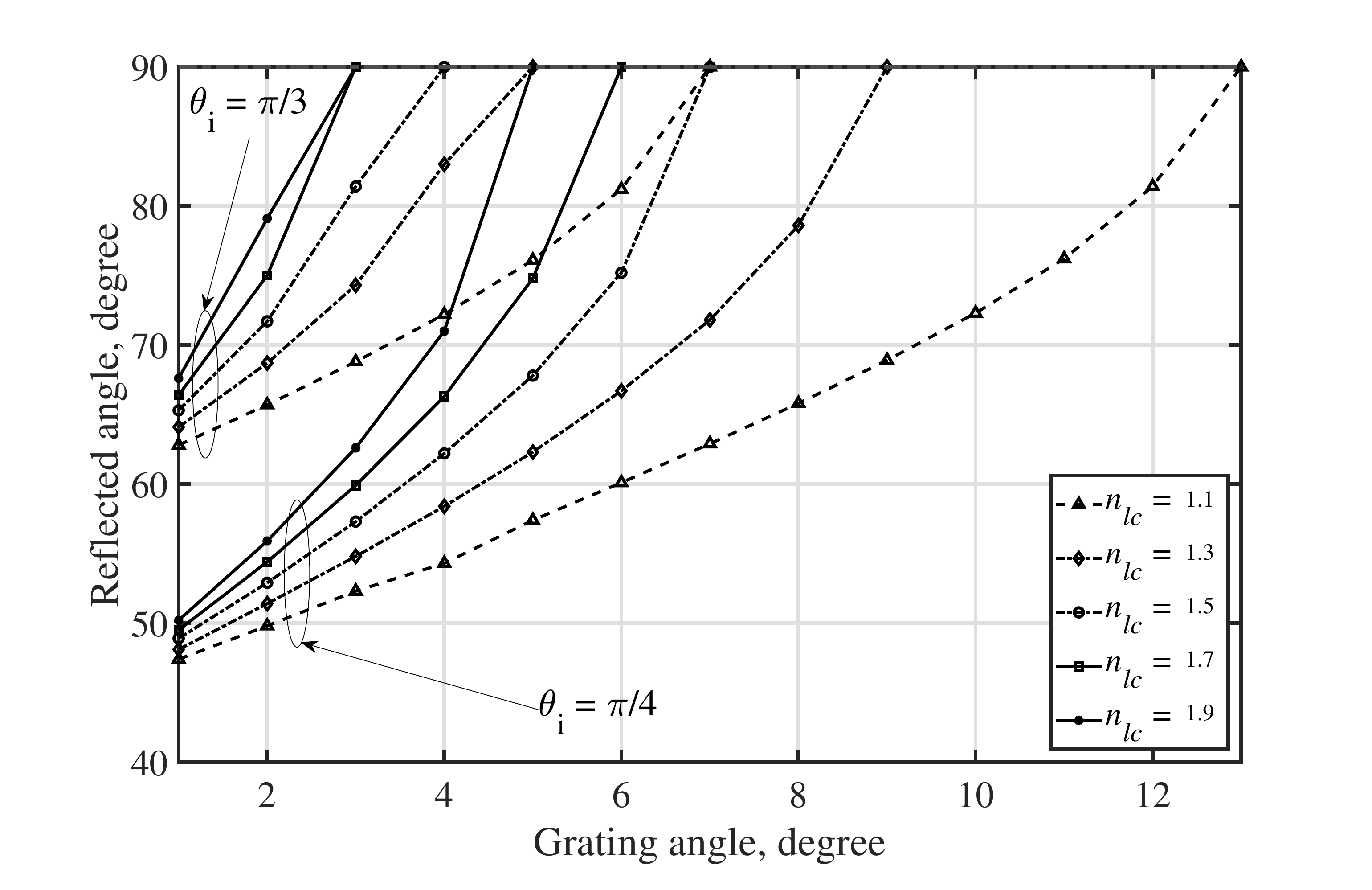}
	\caption{Deflected angles.}
	\label{fig:Angles}
\end{figure} 

\begin{figure}
	\centering
	\includegraphics[width=0.5\textwidth]{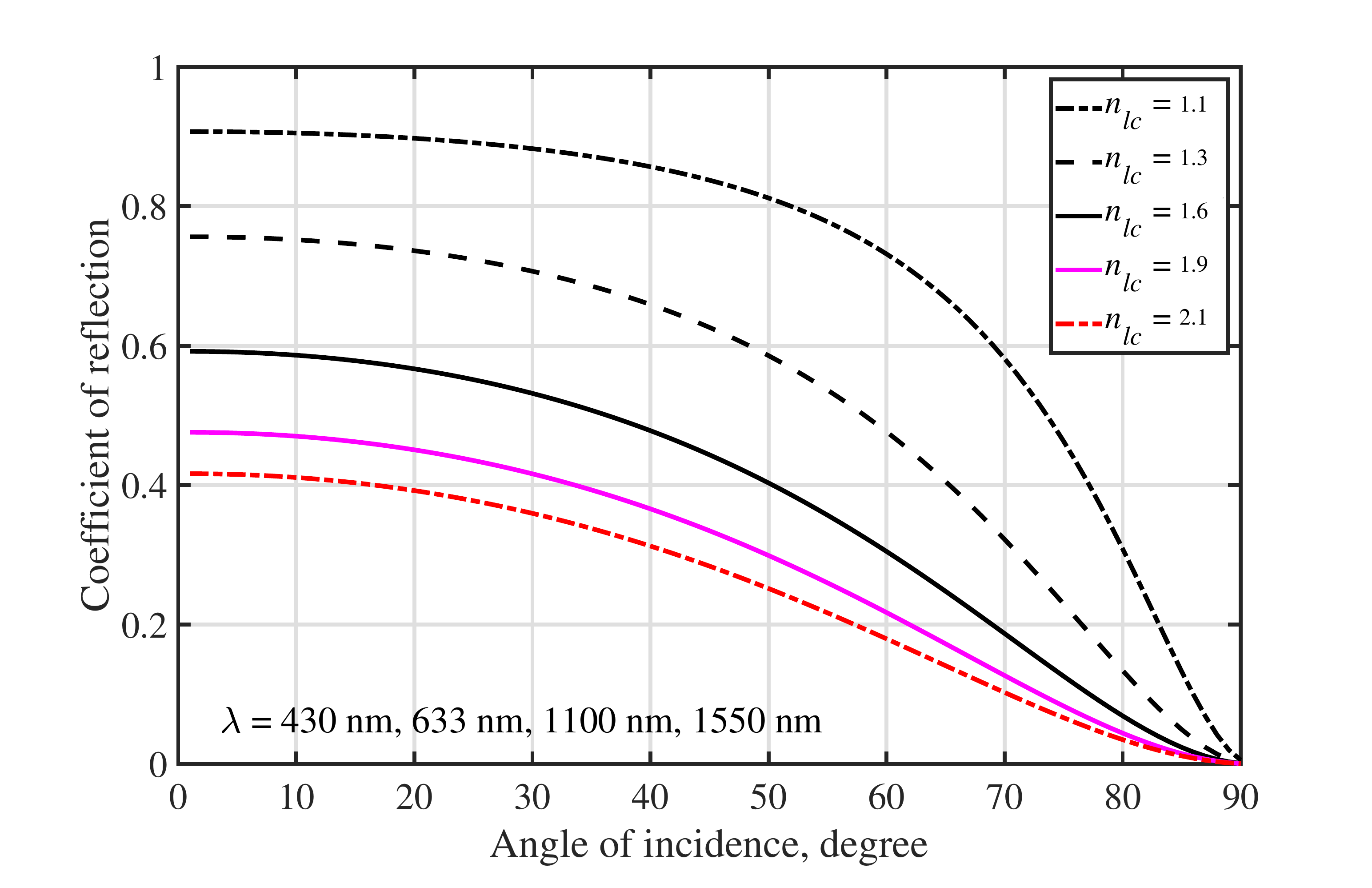}
	\caption{Coefficient of reflection.}
	\label{fig:Coef_Reflec}
\end{figure}

\noindent
\textbf{Geometrical Considerations:} Fig.~\ref{fig:Design} illustrates a cutaway view of a DRIS structure, which is a layered LC-based structure that can modulate and reflect (transmit) light. The reflected angle, $\theta_r$, is given by $\theta_r = \sin^{-1} \left[n \sin \left(2\alpha+\sin^{-1}(n^{-1}\sin \theta_i)\right)\right]$, where $\theta_i$ is the incident angle. The angle $\theta_r$, which is calculated for a single blaze, leads to an infinity of light rays, after scattering occurs when the light hits the structure. In order to avoid this problem, a periodic grating system should be used to create a stronger main lobe. This leads to the following grating equation elaborated from the Maxwell's equations: $ na \left(\sin\varphi_r+\sin\varphi_i\right) = m\lambda $. At the bottom of the LC element, the light is deflected (refracted) with an order $ m = \pm 1, \pm 2, \pm 3, \dots$, and the deflected angle is $\varphi_r = \pm \varphi_i + 2\alpha$. The two key parameters to control $\theta_r$ are $ n $ and $\alpha$. $\alpha$ is determined by the period $ a $ and blase $ b $, which are calculated during the design of the element structure.\footnote{Note that several types of grating have been efficiently used in optics, namely blazed and binary grating, to mention only two \cite{7362130}.} Hence, the tuning is based on varying the birefringence, $\Delta n$. Fig.~\ref{fig:Angles} illustrates the impact of $n$ and $\alpha$ on the orientation of the emerging light. The design considers the value of $\alpha$ that will result in a wider angle of deflection (refraction). Since the emerging light is oriented according to the control bits sequence for any incident angle, it is important to be aware that the reflection coefficient also varies with the incidence angle. Fig.~\ref{fig:Coef_Reflec}, which has been generated based on the matrix of a set of parallel plates (see Fig.~\ref{fig:Design}) given in \eqref{Eq:matrix} \cite[Eq. (1.4-11)]{book3}, clarifies the influence of the incident angle on the reflection (refraction) coefficient,
\begin{equation}
	\textbf{M} = 
	\begin{bmatrix}
		1 & \sum_{j=1}^{\chi}\frac{d_j}{n_j}  \\
		0 & 1  \\
	\end{bmatrix}.
\label{Eq:matrix}
\end{equation}
Fig.~\ref{fig:Coef_Reflec} shows that the reflected light intensity decreases with the angle of incidence and with increasing the refractive index. 
\subsection{Example}
\begin{figure}
	\centering
	\includegraphics[width=0.45\textwidth]{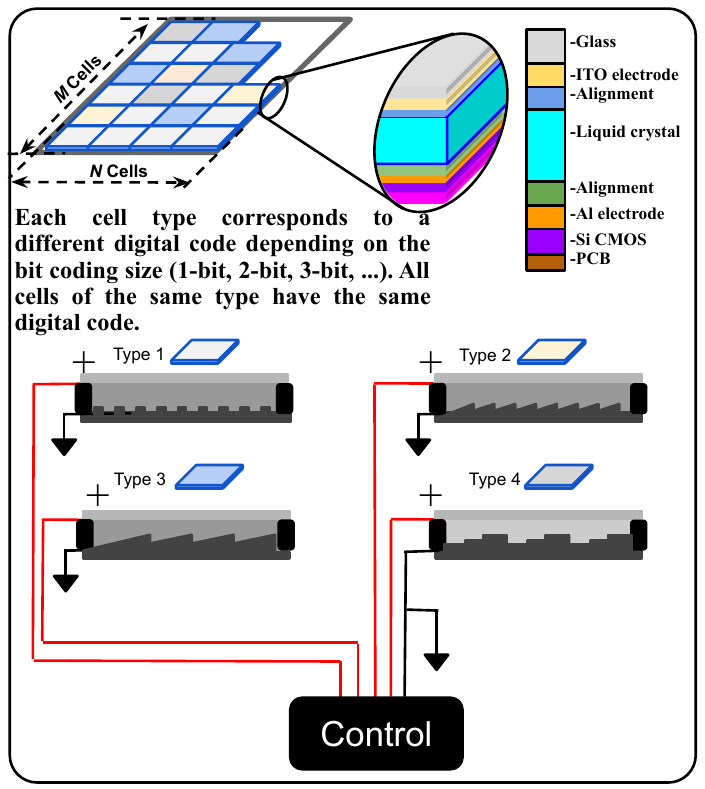}
	\caption{LC-based DRIS ($ X $, 2, $\rho_o$) structure design.}
	\label{fig:Structure}
\end{figure}  
\begin{figure}
	\centering
	\begin{subfigure}[b]{0.5\textwidth}
		\centering
		\includegraphics[width=\textwidth]{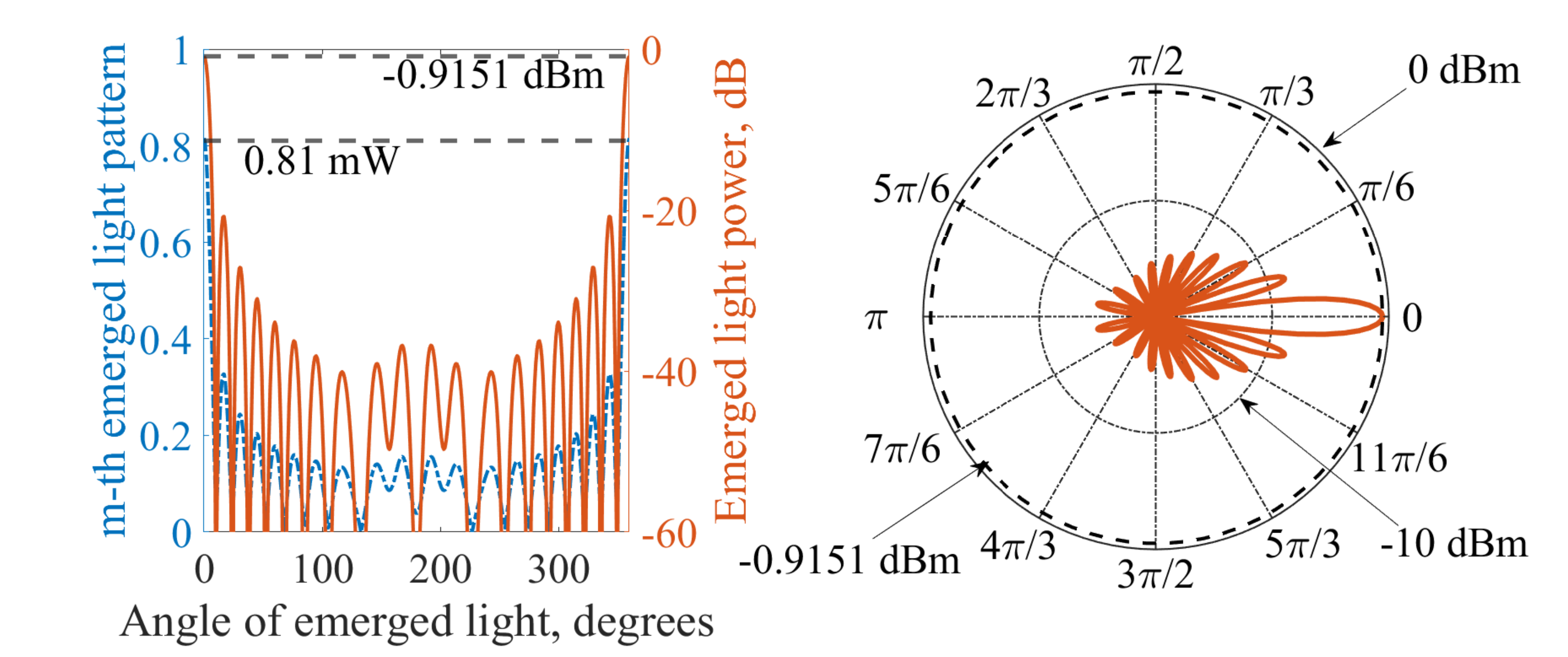}
		\caption{Control sequence 1, "00" ($ \theta_{r1} $ = 0, $ \rho_1 $ = 0.81, $\beta_1$ = 0.9).}
		\label{fig:Seq11}
	\end{subfigure}
	\hfill
	\begin{subfigure}[b]{0.5\textwidth}
		\centering
		\includegraphics[width=\textwidth]{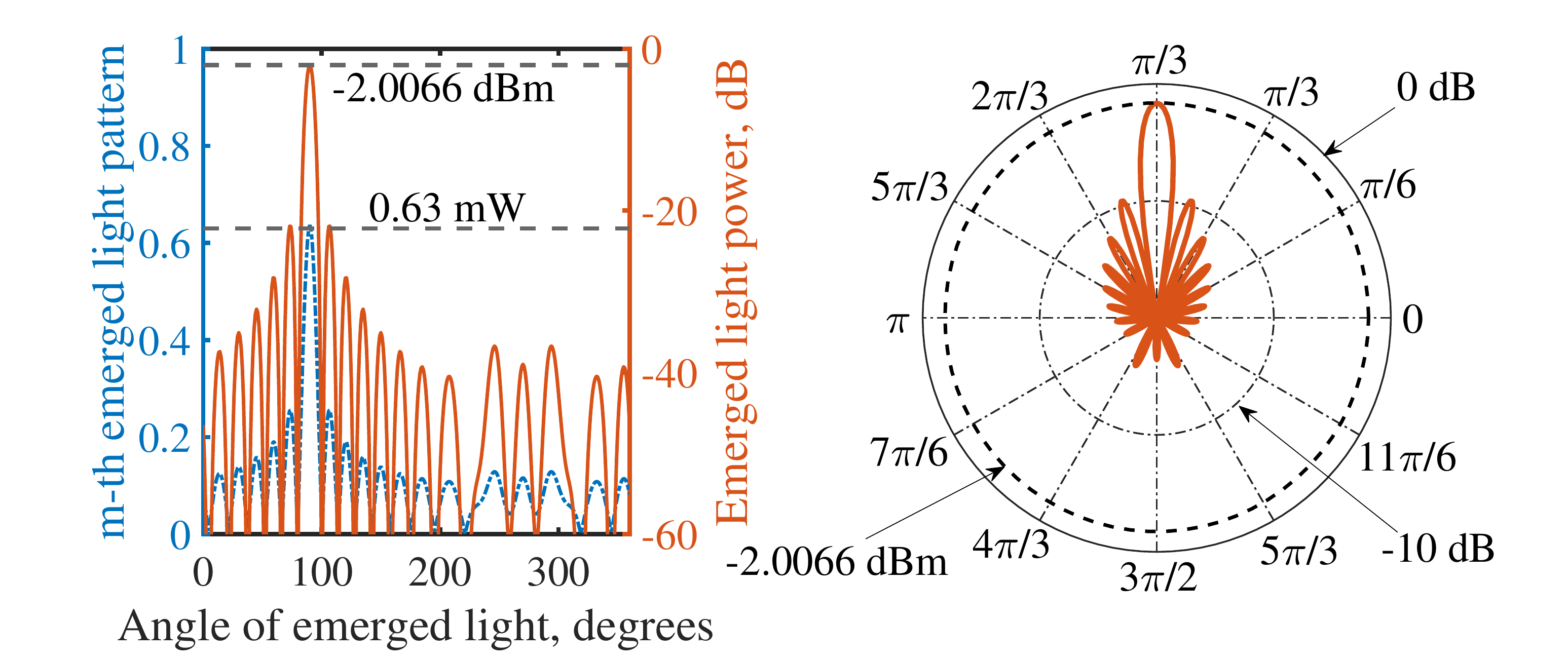}
		\caption{Control sequence 2, "01" ($ \theta_{r2} $ = $ \pi $/2, $ \rho_2 $ = 0.63, $\beta_2$ = 0.7).}
		\label{fig:Seq12}
	\end{subfigure}
	\hfill
		\begin{subfigure}[b]{0.5\textwidth}
		\centering
		\includegraphics[width=\textwidth]{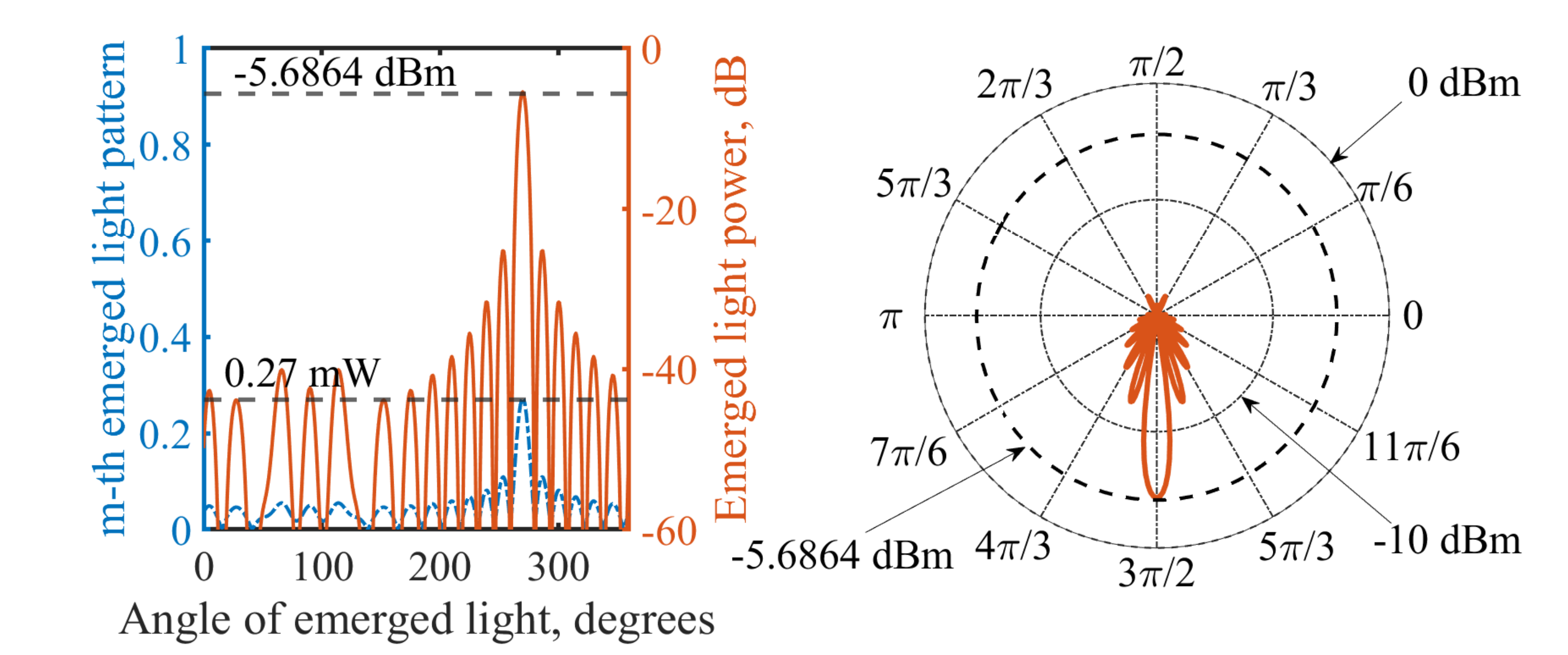}
		\caption{Control sequence 4, "10" ($ \theta_{r4} $ = 3$ \pi $/2, $ \rho_4 $ = 0.27, $\beta_4$ = 0.3).}
		\label{fig:Seq21}
	\end{subfigure}
	\hfill
\begin{subfigure}[b]{0.5\textwidth}
	\centering
	\includegraphics[width=\textwidth]{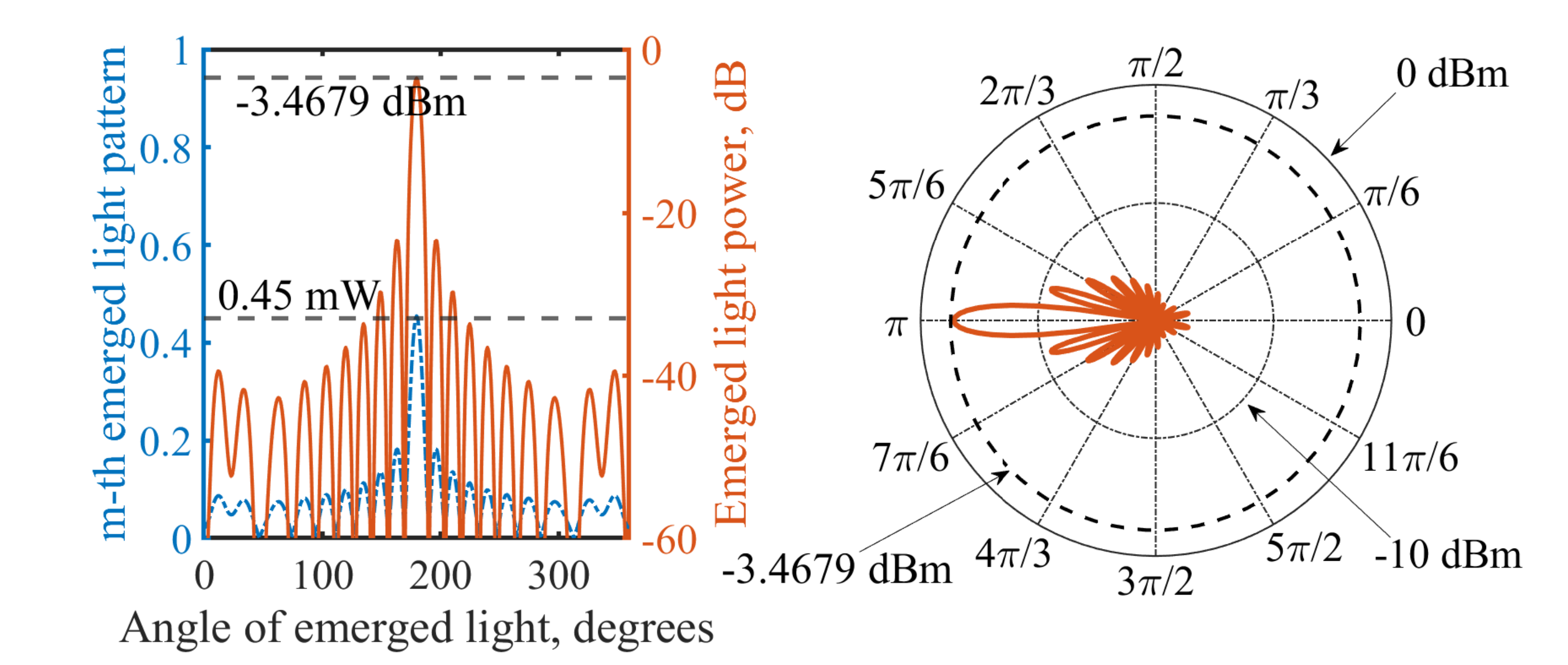}
	\caption{Control sequence 3, "11" ($ \theta_{r3} $ = $ \pi $, $ \rho_3 $ = 0.45, $\beta_3$ = 0.5).}
	\label{fig:Seq22}
\end{subfigure}
	\caption{Example of DRIS (X, 2, 0.9).}
	\label{fig:Seq}
\end{figure}
Fig.~\ref{fig:Structure} illustrates a DRIS ($ X $, 4, 0.9) consisting of a $M \times N$ 2-dimensional device composed of four types of coded elements. The elements consist of miniaturized A4907 panels sandwiched between transparent layers and with thickness $ d$ = 13.34 $\mu $m \cite{zou2021fast}. The critical voltage for the A4907 is 1.3 V, which results in a deviation of 2$\pi$ at 2.6 V under a temperature of 40°C \cite{zou2021fast}. It utilizes a thin-film Si CMOS back-plane as its bottom layer, which acts as a mirror to provide total internal reflection, as well as to generate the necessary temperature for LC operation. Its upper face is designed to create a blazed grating reflection system. When the DRIS is designed as a refractive device, its bottom layer is transparent and the lower face is covered by a blazed grating-type refractive system. The element's upper layer is composed of a glass substrate. The four types of elements ($ u \in$ \{1, 2, 3, 4\}) can reflect or refract the incoming light into four different directions based on the input coding sequence, i.e., "00", "01", "11", and "10", leading to an emerged light oriented at 0, $\pi$/2, $\pi$, and 3$\pi$/2, respectively compared to the normal to the DRIS plan, with $ u $-indexed values 90\%, 70\%, 50\%, 30\%, and reflection coefficients ($\rho_1$, $\rho_2$, $\rho_3$, $\rho_4$) = (0.81, 0.63, 0.45, 0.27), respectively. The emerging lights are oriented regardless of the angle of incidence of the incoming light. These blazed gratings are intended to provide high power to the main reflected or refracted lobe.

Fig.~\ref{fig:Seq} illustrates the element gain, power, and emerged light patterns of the LC-based DRIS structure presented in Fig.~\ref{fig:Structure}. Both the gain and emerged light patterns are shown on the left hand side of the sub-figures. Fig.~\ref{fig:Seq11}, \ref{fig:Seq12}, \ref{fig:Seq21}, and \ref{fig:Seq22} show the elements main beam and the corresponding signal power, which respectively correspond to 0 and -0.9151 dBm, $\pi$/2 and -2.0066 dBm, 3$\pi$/2 and -3.4679 dBm, and $\pi$ and -5.6864 dBm for the four types of element, for 1.3 V, 1.6 V, 1.8 V, and 2.6 V \cite{zou2021fast}.
\section{Ideas for Further Research and Conclusion}
DRIS is a novel concept that deserves more attention due to its numerous benefits. The study of VLC systems with a DRIS module for indoor VLC beam steering needs to be investigated. Applications such as localization, tracking, and positioning, in addition to light fidelity-based applications, among others, may require efficient beamforming in an environment where a strong probability of obstruction of the line-of-sight signal exists. It will be necessary to review all the DRIS applications in OWC systems in order to provide new models for RIS-assisted OWC channels.

The concept of DRIS should be extended to radio frequency (RF)-based wireless networks for beam management in the upcoming B5G network. In RIS-assisted RF networks, DRIS will offer enormous advantages. Therefore, it will be appropriate to study the different materials that can be utilized to develop RF-based DRIS modules. The effect of DRIS on the received signal strength at the users in a RIS-assisted RF network deserves careful investigations.

In this article, we examined the use of DRIS in OWC systems to replace traditional RIS in order to enable DSP methods to physical materials. The basic concepts of coding, digital, programmable, and information metamaterials were defined. Further, we discussed front raw materials suitable for use in optical DRIS. As a consequence, LCs were selected as potential candidates to be used in DRIS for OWC systems. We examined the main parameters of the LCs, namely their refractive index and tilt angle, which are very sensitive to the external electric field. We proposed an example of LC-based DRIS design using the A4907 with four element types, which is triggered at 1.3 V and is controlled with 1.3 V, 1.6 V, 1.8 V, and 2.6 V to generate beam at 0 and -0.9151 dBm, $\pi$/2 and -2.0066 dBm, 3$\pi$/2 and -3.4679 dBm, and $\pi$ and -5.6864 dBm, respectively.
	\ifCLASSOPTIONcaptionsoff
	\fi
\bibliographystyle{IEEEtran}
\bibliography{Digital}
\end{document}